\documentclass[aps,prl,twocolumn,floatfix,superscriptaddress]{revtex4-1}

\usepackage{amsmath}
\usepackage{amssymb}
\usepackage{times}
\usepackage{braket}
\usepackage[pdftex]{graphicx}
\usepackage{color}
\usepackage{placeins}
\usepackage{blindtext}

\renewcommand{\vec}[1]{\boldsymbol{#1}}
\newcommand{\change}[1]{\textcolor{black}{#1}}

\begin{document}

\title{Magnetic bimerons as skyrmion analogues in in-plane magnets}

\author{B{\"o}rge G{\"o}bel}
\email[Corresponding author: ]{bgoebel@mpi-halle.mpg.de}
\affiliation{Max-Planck-Institut f\"ur Mikrostrukturphysik, D-06120 Halle (Saale), Germany}

\author{Alexander Mook}
\affiliation{Institut f\"ur Physik, Martin-Luther-Universit\"at Halle-Wittenberg, D-06099 Halle (Saale), Germany}

\author{J\"urgen Henk}
\affiliation{Institut f\"ur Physik, Martin-Luther-Universit\"at Halle-Wittenberg, D-06099 Halle (Saale), Germany}

\author{Ingrid Mertig}
\affiliation{Max-Planck-Institut f\"ur Mikrostrukturphysik, D-06120 Halle (Saale), Germany}
\affiliation{Institut f\"ur Physik, Martin-Luther-Universit\"at Halle-Wittenberg, D-06099 Halle (Saale), Germany}

\author{Oleg A. Tretiakov}
\affiliation{School of Physics, The University of New South Wales, Sydney 2052, Australia}
\affiliation{Institute for Materials Research and Center for Science and Innovation in Spintronics, Tohoku University, Sendai 980-8577, Japan}
\affiliation{National University of Science and Technology MISiS, Moscow 119049, Russia}

\date{\today}

\begin{abstract}
A magnetic bimeron is a pair of two merons and can be understood as the in-plane magnetized version of a skyrmion. Here we theoretically predict the existence of single magnetic bimerons as well as bimeron crystals, and compare the emergent electrodynamics of bimerons with their skyrmion analogues.  We show that bimeron crystals can be stabilized in frustrated magnets and analyze what crystal structure can stabilize bimerons or bimeron crystals via the Dzyaloshinskii-Moriya interaction. We point out that bimeron crystals, in contrast to skyrmion crystals, allow for the detection of a pure topological Hall effect. By means of micromagnetic simulations, we show that bimerons can be used as bits of information in in-plane magnetized racetrack devices, where they allow for current-driven motion for torque orientations that leave skyrmions in out-of-plane magnets stationary.
\end{abstract}

\maketitle

\begin{figure*}
  \centering
  \includegraphics[width=\textwidth]{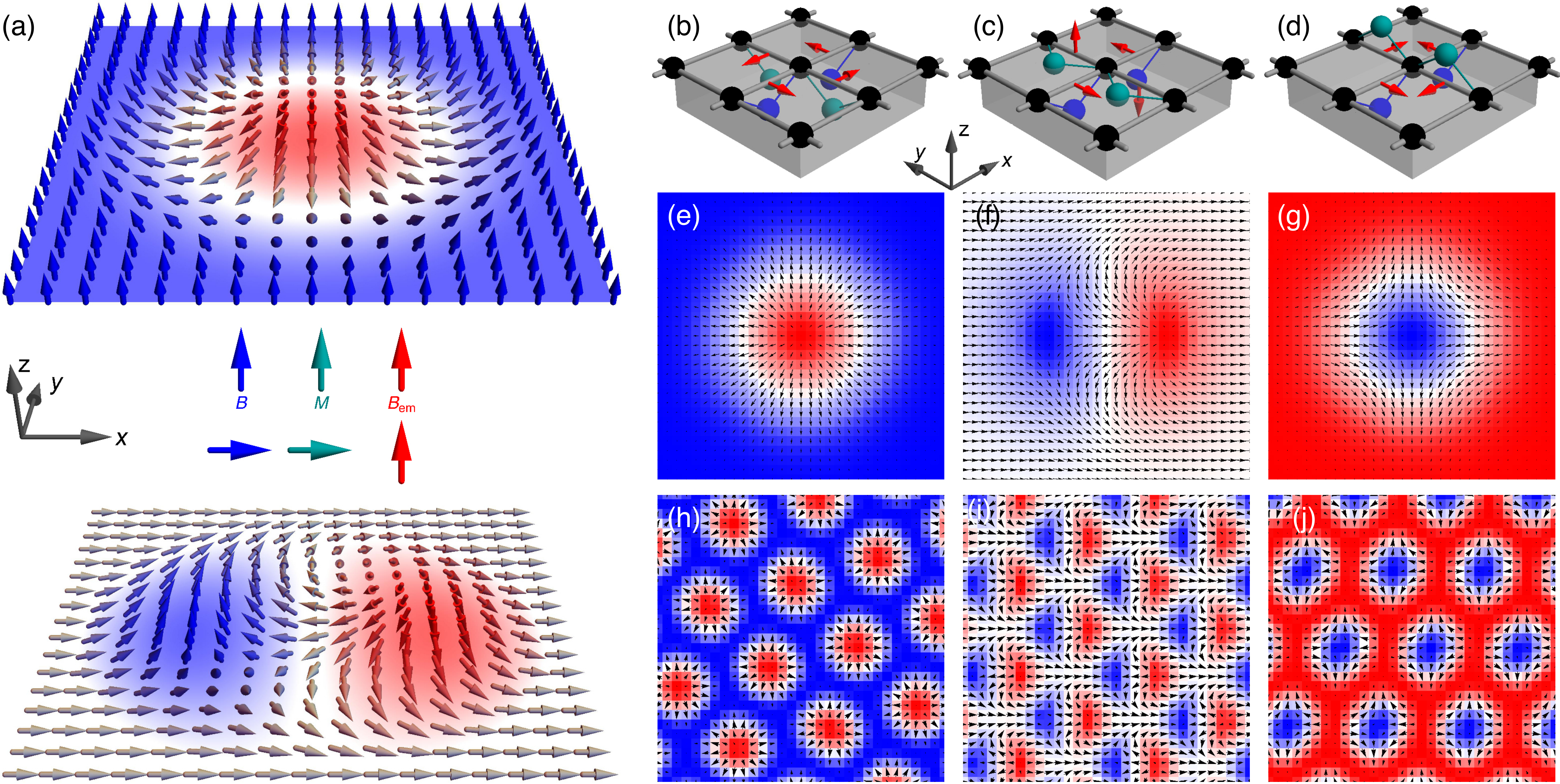}  
  \caption{Magnetic textures. (a) If the magnetic moments of a magnetic skyrmion (top) are rotated by $90^\circ$ around the $y$ axis the resulting texture is a magnetic bimeron (bottom), which requires a stabilizing magnetic field $\vec{B}$ and has net magnetization $\vec{M}$, both rotated in the same way. The emergent field $\vec{B}_\mathrm{em}$ remains out-of plane. (b-d) show the necessary geometry of magnetic atoms (black) and heavy metal atoms (blue, green) to generate the DMI vectors (red) which stabilize (e-g) isolated skyrmions, bimerons and antiskyrmions, respectively. (h-j) show periodic arrays. The color in the magnetic textures indicates the out-of plane component of the magnetization (blue: positive, red: negative). Panel (a) shows schematic representations. (e-g) are relaxed metastable textures from propagating the LLG equation in a ferromagnetic background at zero temperature for the parameters: $D=J/3$, $|K|=|B|=J/15$ (here $J$ is the nearest-neighbor exchange constant) and Gilbert damping $\alpha=0.01$. (h-j) are results of Monte Carlo simulations at zero temperature, with $D=J$, $|B|=|K|=0.3\,J$. Magnetic field $\vec{B}$ and easy-axis anisotropy $\vec{K}$ point along $z$ for the skyrmion, $x$ for the bimeron and $-z$ for the antiskyrmion; the corresponding DMI vectors have been used.}
  \label{fig:dmi}
\end{figure*}

Over the last years magnetic skyrmions [Fig.~\ref{fig:dmi}(a) top]~\cite{skyrme1962unified,bogdanov1989thermodynamically,bogdanov1994thermodynamically, rossler2006spontaneous,muhlbauer2009skyrmion,nagaosa2013topological} have attracted immense research interest, as these small spin textures $\vec{m}(\vec{r})$ possess strong stability, characterized by a topological charge $N_\mathrm{Sk}=\pm 1$.
Skyrmions offer a topological contribution to the Hall effect~\cite{bruno2004topological,neubauer2009topological,schulz2012emergent,kanazawa2011large, lee2009unusual,li2013robust,hamamoto2015quantized,lado2015quantum,
gobel2017THEskyrmion,gobel2017QHE,ndiaye2017topological,yin2015topological}, commonly measured in skyrmion crystals, and can be stabilized as individual quasiparticles in collinear ferromagnets. They can be driven by currents in thin films~\cite{jonietz2010spin,sampaio2013nucleation,nagaosa2013topological,zang2011dynamics, iwasaki2013current,jiang2017direct,litzius2017skyrmion, tomasello2014strategy,gobel2019overcoming} allowing for spintronics applicability. The stabilizing interaction in most systems is the Dzyaloshinskii-Moriya interaction (DMI)~\cite{dzyaloshinsky1958thermodynamic,moriya1960anisotropic}, while theoretical simulations also point to other stabilizing mechanisms, e.\,g. frustrated exchange interactions~\cite{okubo2012multiple,leonov2015multiply}. \change{Textures with a half-integer topological charge, like merons and antimerons (or vortices and antivortices), have also been subject of intense research~\cite{kosevich1990magnetic,mertens2000dynamics,sheka2008electromagnetic}.}

Magnetic bimerons~\change{\cite{kharkov2017bound,heo2016switching, zhang2015magnetic,komineas2007rotating}} [Fig.\ref{fig:dmi}(a) bottom] are the combination of two merons [red and blue] and can be understood as in-plane magnetized versions of magnetic skyrmions~\footnote{Note, that recently the term 'bimeron' has also been used for elongated skyrmions~\cite{ezawa2011compact,du2013field,silva2014emergence}, instead of the original object, found in dual layer two-dimensional electron gases and quantum Hall systems~\cite{moon1995spontaneous,yang1995charged,brey1996charged,bourassa2006pseudospin,
cote2010orbital}. Throughout this \change{Paper} we always refer to the latter.}. Instead of the out-of-plane component of the magnetization it is an in-plane component which is radial symmetric about the quasiparticle's center; being aligned with the saturation magnetization of the ferromagnet at the outer region of the bimeron and pointing into the opposite direction in the center. Recently, Kharkov \textit{et al.} showed that isolated bimerons can be stabilized in an easy-plane magnet by frustrated exchange interactions~\cite{kharkov2017bound}. In DMI dominated systems (as is the case for all experimentally known skyrmion-host materials) bimerons have only been shown to exist as unstable transition states~\cite{heo2016switching,zhang2015magnetic}. 

In this \change{Rapid Communication}, we show that bimerons in frustrated magnets can also be stabilized in an array, the bimeron crystal. Furthermore, we propose a structural configuration that allows for DMI stabilizing isolated bimerons and bimeron crystals. We compare fundamental properties of skyrmions and bimerons and find that both show the same topological Hall effect, whereas the bimeron allows for a \emph{pure} detection, that is without superposition of the anomalous and ordinary Hall effects. Elaborating on the spintronics applicability of bimerons in in-plane racetrack memory devices, we find that bimerons can be driven by spin currents, similar to skyrmions. However, they extend the class of materials and spin-torque configurations for building spintronics devices. 

\paragraph{Stabilization of bimerons and bimeron crystals.}

A bimeron [see Fig.~\ref{fig:dmi}(a) bottom] \change{(or a vortex-antivortex pair)} consists of two subtextures: a meron and an antimeron \change{(or a vortex and an antivortex)}, with mutually reversed $z$ components of the magnetic moments $\{\vec{m}_i\}$. Still, the bimeron itself is the quasiparticle in in-plane magnets, since merons and antimerons can not exist individually in a ferromagnet. The topological charge density 
\begin{align}
n_\mathrm{Sk}(\vec{r})  = \frac{1}{4\pi} \vec{m}(\vec{r}) \cdot \left[ \frac{\partial \vec{m}(\vec{r})}{\partial x}  \times  \frac{\partial \vec{m}(\vec{r})}{\partial y}  \right]\label{eq:nsk}
\end{align}
is distributed radially symmetrically around the center of the bimeron and integrates to $N_\mathrm{Sk}=\pm 1$; meron and antimeron carry a topological charge of $\pm 1/2$ each~\cite{Tretiakov2007}.

The recurring idea of this \change{Paper} is a geometric comparison of skyrmions, bimerons, and antiskyrmions: all three magnetic textures are related by a rotation of each spin around an in-plane axis (in this \change{Paper} always $y$). A bimeron is constructed by rotating each spin of a skyrmion by $90^\circ$ [cf. Fig.~\ref{fig:dmi}(a)]; for an antiskyrmion the spins have to be rotated by another $90^\circ$. 

To find stable bimerons or bimeron crystals one can therefore start from any system that stabilizes skyrmions and rotate every vectorial term in the Hamiltonian. The most effortless approach is to consider skyrmions stabilized by frustrated exchange $-J_{ij}\vec{m}_i\cdot\vec{m}_j$. If the \emph{scalar} constants $J_{ij}$ for nearest and next-nearest neighbor interactions have opposite signs the ground state of the system \change{can be} a spin-spiral phase~\cite{okubo2012multiple}. When an external magnetic field $\vec{B}$ and easy-axis anisotropy $\vec{K}$ are present pointing out-of-plane, skyrmions and skyrmion crystals may be stabilized. 

Following this idea, bimerons and bimeron crystals are stabilized in a system where both $\vec{B}$ and $\vec{K}$ are rotated in-plane [cf. Fig.~\ref{fig:dmi}(a)]. Then, the Hamiltonian
\begin{align*}
H=&-\frac{1}{2}\sum_{i,j}J_{ij}\vec{m}_i\cdot\vec{m}_j-\sum_i\vec{B}\cdot\vec{m}_i\\
&-\frac{1}{2}\sum_i\sum_{A\in\{x,y,z\}}K_A\left(m_i^A\right)^2\label{eq:ham}
\end{align*}
gives the same energy as for the skyrmion phase before the rotation. The results of Monte Carlo simulations confirming this finding are presented in the Supplemental Material~\cite{SupplementalMaterial}. The analogy of the two systems does also hold for other typical phases: For low fields we find a spin-spiral state, for medium fields the bimeron crystal and for high fields the system is fully magnetized. At the transition we find isolated bimerons in an in-plane magnetized  background.

To illustrate the geometric equivalence of bimeron and skyrmion we used an easy-axis anisotropy along an in-plane direction, even though such quantity is commonly small. Our results also hold for systems without anisotropy or with an easy-plane anisotropy (as in Ref.~\onlinecite{kharkov2017bound}, where isolated meta-stable bimerons have been considered), since the applied magnetic field makes the two in-plane directions inequivalent (see~\cite{SupplementalMaterial}).

At the present state of research almost all experimentally detected skyrmions are generated by the Dzyaloshinskii-Moriya interaction (DMI)~\cite{dzyaloshinsky1958thermodynamic,moriya1960anisotropic}
\begin{align}
H_\mathrm{DMI}=\frac{1}{2}\sum_{i,j}\vec{D}_{ij}\cdot(\vec{m}_i\times\vec{m}_j).
\end{align} 
It is a relativistic energy contribution due to spin-orbit coupling and broken inversion symmetry. The DMI vectors $\vec{D}_{ij}$ obey Moriya's symmetry rules~\cite{moriya1960anisotropic} and can be estimated from the Levy-Fert rule~\cite{fert1980role}; $\vec{D}_{ij}$ points into the direction $\vec{r}_{i\rightarrow j}\times \vec{r}_{i\rightarrow \mathrm{HM}}$, i.\,e., it is perpendicular to the plane of the two lattice sites $i,j$ and the nearest heavy-metal atom (HM). Similar to the frustrated exchange interactions the DMI leads to spin canting, but since it is vectorial it strictly dictates the type of magnetic texture: Skyrmions can not be turned into bimerons by rotating $\vec{B}$ and $\vec{K}$ only. \change{The} $\vec{D}_{ij}$ have to be adjusted as well~\footnote{\change{Note, that in magnets with in-plane anisotropy and conventional interfacial DMI [Fig.~\ref{fig:dmi}(b)] asymmetric skyrmions are stabilized~\cite{leonov2017asymmetric}. These objects are intermediate states between skyrmions and bimerons with non-vanishing in-plane and out-of-plane net magnetization.}}.

At interfaces of layered systems [Fig.~\ref{fig:dmi}(b)] heavy-metal atoms (green and blue) induce DMI vectors between neighboring magnetic atoms (black). Typically, the DMI vectors form a toroidal arrangement and produce N\'{e}el skyrmions (e) or N\'{e}el skyrmion crystals (h). Rotating the HM atoms around the bond in $y$ direction, the $\vec{D}_{ij}$ are rotated in the same way according to the Levy-Fert rule. If now external field and anisotropy are oriented along the $x$ direction, as in the frustrated exchange case, bimerons or bimeron crystals are stabilized for the same parameters (in magnitude) as for the skyrmion phase, see Fig.~\ref{fig:dmi}(f) and (i). This approach is confirmed by Monte Carlo simulations and atomistic simulations of the Landau-Lifshitz-Gilbert equation (LLG)~\cite{landau1935theory,gilbert2004phenomenological} (see Supplemental Material~\cite{SupplementalMaterial}). To complete this picture, we point out that for the stabilization of antiskyrmions (g, j) the indicated HM atoms (green) have to be rotated another $90^\circ$ around the bond in $y$ direction (d) --- a configuration recently found experimentally~\cite{nayak2017magnetic} in the Heusler alloy Mn$_{1.4}$Pt$_{0.9}$Pd$_{0.1}$Sn. The corresponding DMI is called `anisotropic'~\cite{Gungordu2016,huang2017stabilization,hoffmann2017antiskyrmions}.

Summarizing up to this point we predict the existence of isolated bimerons and bimeron crystals by frustrated exchange and DMI. 
Next, we discuss implications of the in-plane magnetized bulk systems and thin films with bimerons for electronic properties and spintronic applications.

\paragraph{Pure topological Hall effect of electrons.}

\begin{figure}
  \centering
  \includegraphics[width=\columnwidth]{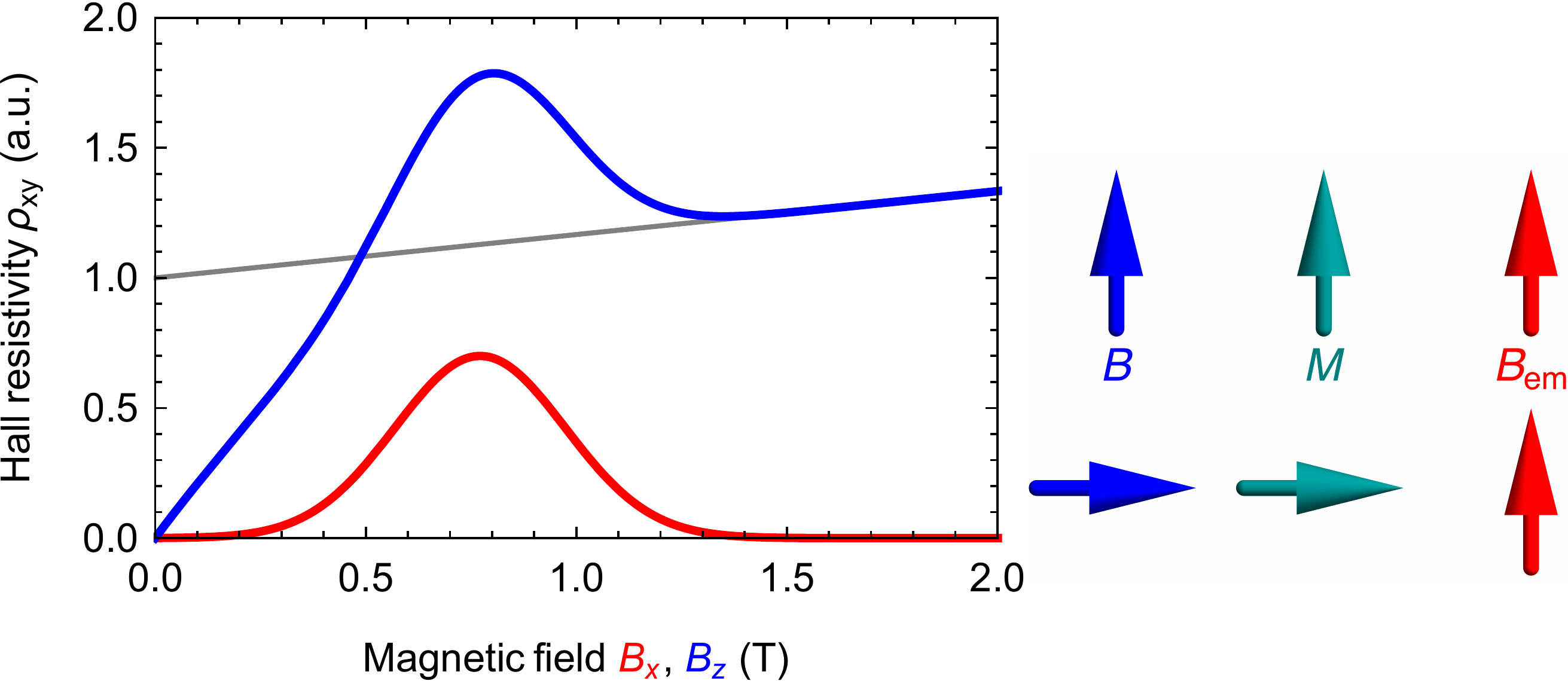}
  \caption{Hall resistivity (schematic). We compare the signal for a bimeron (red) and a skyrmion (blue). While the signal is purely of topological origin in the bimeron phase (here stable for magnetic fields in $x$ direction between $0.5\,\mathrm{T}$ and $1\,\mathrm{T}$), it is superimposed by the ordinary Hall (cf. slope of the gray curve) and the anomalous Hall effect (offset of the gray curve) for the skyrmion (here the stabilizing field is applied along $z$).}
  \label{fig:the}
\end{figure}

When an electric field $\vec{E}$ is applied to a metal, a current $\vec{j}$ flows according to Ohm's law $\vec{E}=\rho\vec{j}$. For a skyrmion crystal the transverse element of the resistivity tensor
\begin{align}
\rho_{xy}=\rho_{xy}^\mathrm{HE}+\rho_{xy}^\mathrm{AHE}+\rho_{xy}^\mathrm{THE}\label{eq:hall}
\end{align} 
is decomposed into an ordinary Hall contribution~\cite{hall1879new} due to an external magnetic field $\rho_{xy}^\mathrm{HE}\propto B_z$, an anomalous Hall contribution~\cite{nagaosa2010anomalous} due to spin-orbit coupling and a net magnetization $\rho_{xy}^\mathrm{AHE}\propto M_z$~\footnote{\change{Over the recent years an additional contribution to the Hall resistivity $\rho_{xy}$ has been predicted which is neither proportional to $M$ nor $N_\mathrm{sk}$; it is only determined by the SOC~\cite{chen2014anomalous}. 
Even though this effect can not be excluded, all Hall measurements in skyrmion crystals are well approximated by Eq.~(\ref{eq:hall}) to the best of our knowledge.}}, and a topological Hall contribution due to the local topological charge density [Eq.~\eqref{eq:nsk}] that acts like an emergent field $\rho_{xy}^\mathrm{THE}\propto \braket{B_{\mathrm{em},z}}\propto N_\mathrm{Sk}$. For skyrmions $\vec{B}$, $\vec{M}$, and $\braket{\vec{B}_\mathrm{em}}$ point along the $z$ direction.

For a bimeron the spin rotation renders the $z$ component of magnetic quantities zero, $B_z=M_z=0$ [cf. Fig.~\ref{fig:dmi}(a)], but since $n_\mathrm{Sk}$ is invariant under global spin rotation $B_{\mathrm{em},z}$ remains. For this reason \emph{only} the topological Hall effect emerges in a sample with bimerons (see Fig.~\ref{fig:the}). This hallmark for real-space topology can be detected in an isolated manner making bimeron crystals a playground for investigating fundamental physics.
In the Supplemental Material~\cite{SupplementalMaterial} we numerically validate the equivalence of the topological Hall effect for skyrmion and bimeron crystals following Refs.~\onlinecite{hamamoto2015quantized,gobel2017THEskyrmion,gobel2017QHE, ndiaye2017topological,yin2015topological,gobel2018family}, in which the energy-dependent conductivity is discussed.
\paragraph{Current-driven motion in thin film.}

\begin{figure*}
  \centering
  \includegraphics[width=\textwidth]{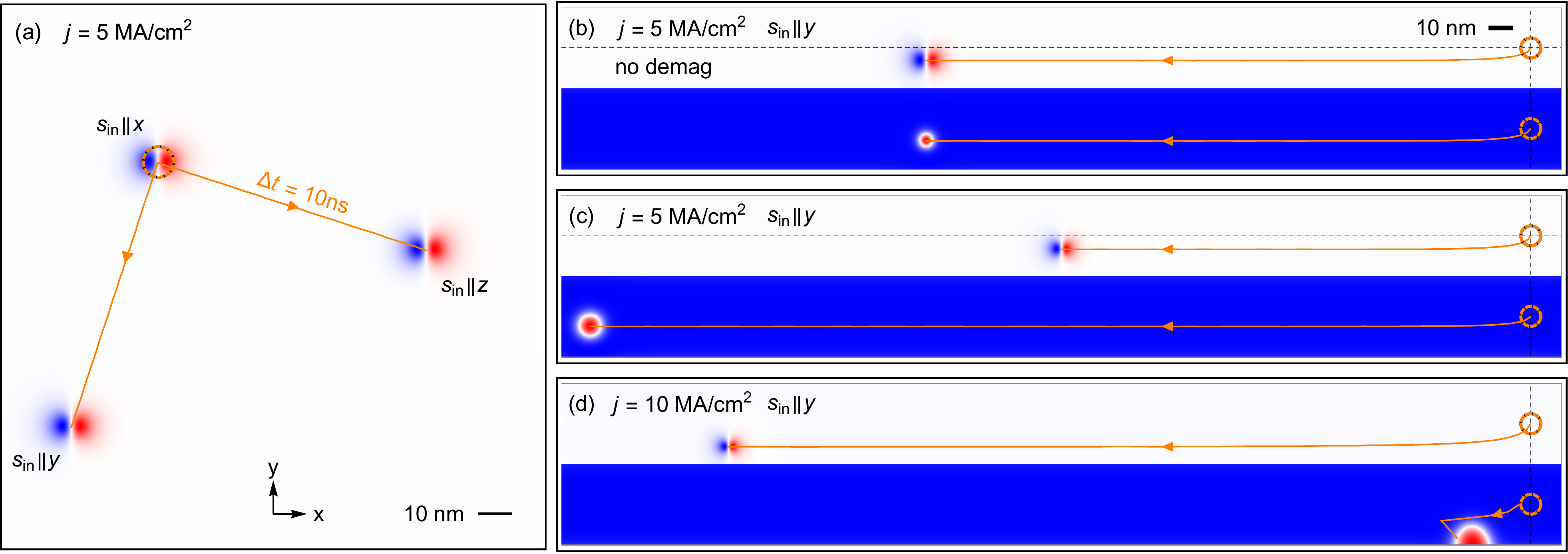}
  \caption{Current-driven motion of magnetic bimerons and skyrmions. (a) Superposition of three results of micromagnetic simulations of a bimeron in an in-plane magnet with periodic boundary conditions starting from the orange circle (top left). The bimerons are driven by SOT type (i) (injected spins $\vec{s}_\mathrm{in}\parallel\vec{x},\vec{y},\vec{z}$ as indicated). (b) Bimeron (top) and skyrmion (bottom) in a racetrack geometry driven by SOT type (ii) (current along the track $j_x=5\mathrm{MA}/\mathrm{cm}^2$, injected spins $\vec{s}_\mathrm{in}\parallel\vec{y}$). The demagnetization field is set to zero. (c) Like (b) but with demagnetization field. (d) Like (c) but at double the current density. In all panels the trajectory (orange) of the center of the bimeron or skyrmion is indicated; the figures show the results after $10\,\mathrm{ns}$ propagation time, except for (d bottom), where the skyrmion already annihilates after $0.5\,\mathrm{ns}$.}
  \label{fig:micromagnetism}
\end{figure*}

\change{In Ref.~\onlinecite{komineas2007rotating} the rotation of an annihilating vortex-antivortex pair in in-plane magnets without DMI has been analyzed. We are now able to discuss the current-driven propagation of \emph{metastable} bimerons.}
In the following, we show that bimerons can be utilized as topologically protected information carriers in in-plane magnetized thin films and discuss similarities and differences to skyrmion racetrack devices~\cite{sampaio2013nucleation,fert2013skyrmions,parkin2004shiftable, parkin2008magnetic,parkin2015memory}. 

In the spin-transfer toque (STT) scenario~\cite{sampaio2013nucleation} a current $\vec{j}$ of spin-polarized electrons is applied along the ferromagnet. Since the electron spin at site $i$ is given by the magnetic texture itself, the torque is rotated in the same way as the magnetization, which leads to identical motion of bimerons and skyrmions under STT.

A more efficient way to drive skyrmions is the spin-orbit torque (SOT) scenario~\cite{sampaio2013nucleation}: spins are injected perpendicularly to the ferromagnetic film, via (i) a spin-polarized current traversing a second ferromagnetic layer with a distinct magnetization $\vec{s}_\mathrm{in}$ or via (ii) a charge current in an adjacent heavy-metal layer, which is transformed into a spin current by the spin Hall effect ($\vec{s}_\mathrm{in}\parallel\vec{y}$ in cubic systems). 
The perpendicularly injected spins are independent of the magnetization in the actual racetrack layer, and large torques can be generated.

The motion of magnetic textures in nanostructured samples is simulated within the micromagnetic approach, that models magnetic textures on a larger length scale compared to the atomistic simulations presented in Fig.~\ref{fig:dmi}. 
We solve the LLG equation (see~\cite{SupplementalMaterial}) for each micromagnetic moment $\vec{m}_i$ with the in-plane spin torque~\cite{slonczewski1996current} proportional to 
\begin{align}
\frac{jP}{d_zM_s}[(\vec{m}_i\times\vec{s}_\mathrm{in})\times\vec{m}_i],
\end{align}
where $d_z$ is the layer thickness, $M_s$ is the saturation magnetization, and $P$ is the spin polarization of a perpendicular current $j$ for (i) or spin Hall angle for (ii).
For comparability the parameters of Co/Pt are taken from Ref.~\onlinecite{sampaio2013nucleation} (they are specified in the Supplemental Material~\cite{SupplementalMaterial}).
The DMI that stabilizes bimerons is derived from the vectors of Fig.~\ref{fig:dmi}(c) 
\begin{align}
\epsilon_\mathrm{DMI}=D\left(m_z\frac{\partial m_x}{\partial x}-m_x\frac{\partial m_z}{\partial x}+m_x\frac{\partial m_y}{\partial y}-m_y\frac{\partial m_x}{\partial y}\right)\notag
\end{align} 
and was implemented in Mumax3~\cite{vansteenkiste2011mumax,vansteenkiste2014design}.

For the SOT scenario (i) skyrmions in a $z$ magnetized ferromagnet can be driven by injected spins $\vec{s}_\mathrm{in}\perp \vec{z}$. Due to the global rotation of spins a bimeron in an $x$ magnetized ferromagnet can be driven by spins $\vec{s}_\mathrm{in}\perp \vec{x}$ and remain stationary for $\vec{s}_\mathrm{in}\parallel \vec{x}$, see Fig.~\ref{fig:micromagnetism}(a).

Towards utilization in a racetrack device the current-driven motion ($\vec{j}\parallel \vec{x}$) is the most relevant aspect of SOTs. Using a cubic heavy metal material for scenario (ii) (e.\,g. Pt), i.\,e., $\vec{s}_\mathrm{in}\parallel\vec{y}$, skyrmions and bimerons are propelled equally in a system with their favoring easy-axis anisotropy ($K_z>0$ for the skyrmion and $K_x>0$ for the bimeron) and DMI [Fig.~\ref{fig:dmi}(b) for the skyrmion and Fig.~\ref{fig:dmi}(c) for the bimeron], as long as the demagnetization field is neglected, cf. Fig.~\ref{fig:micromagnetism}(b). \change{In this case both quasiparticles experience the same forces and behave equally under the influence of temperature.}

The demagnetization field acts effectively as an inhomogeneous in-plane magnetic field for \emph{both} textures, leading to an increase of the skyrmion size and a decrease of the bimeron size. Consequently the skyrmion velocity is larger than that of the bimeron [cf. Fig.~\ref{fig:micromagnetism}(c) and see Supplement~\cite{SupplementalMaterial} for a complementary Thiele equation analysis]. Still, the bimeron can reach similar velocities as the skyrmion since the bimeron allows for larger applied currents densities. While a bimeron is still stable at $j=10\,\mathrm{MA}/\mathrm{cm}^2$, the skyrmion is already annihilated at the edge of the racetrack for $j\gtrsim 6.25\,\mathrm{MA}/\mathrm{cm}^2$. Both quasiparticles can reach velocities of around $50\,\mathrm{m}/\mathrm{s}$ \change{although the skyrmion moves more efficiently for the presented parameters}. In the Supplemental Material~\cite{SupplementalMaterial} we show that current-driven motion is also possible for a material with an easy-plane anisotropy (${K_z < 0}$), when a magnetic field is applied in-plane to generate a preferred direction.
\paragraph{Conclusion and perspective.}
In this \change{Rapid Communication} we have demonstrated how to generate isolated bimerons and bimeron crystals via DMI and frustrated exchange interactions. Since the magnetic moments of a bimeron are merely rotated moments of a skyrmion, the topological properties of the two objects are unchanged and the topological Hall effects due to both of them are identical. Nevertheless, the fact that all magnetic quantities (net magnetization and stabilizing field) are rotated, while the emergent field is not, allows for the pure and therefore unambiguous detection of the topological Hall effect in bimeron systems, and for the development of future spintronic devices based on this effect.

We have shown that magnetic materials with in-plane magnetization can be used to build racetrack storage devices with magnetic bimerons as carriers of information. In these materials, the current-induced dynamics of bimerons can be accomplished similarly to that of skyrmions in conventional racetracks. Furthermore, in-plane ferromagnets allow us to use different orientations of injected spins for the propulsion of bimerons as well as for their generation (for ${\vec{s}_\mathrm{in}\parallel\vec{x}}$ in analogy to Refs.~\onlinecite{sampaio2013nucleation,romming2013writing}). A technological advantage of these materials is the stackability of the quasi-one-dimensional racetracks, since the dipolar energy of two in-plane magnets is smaller than that of two out-of-plane magnets.
The smaller stray fields in a bimeron-based racetrack allow for a denser array of tracks in three dimensions and thus a higher storage density.

The established analogy between skyrmions and bimerons can be carried over to all types of skyrmion-related spin textures such as higher-order skyrmions~\cite{ozawa2017zero,leonov2015multiply}, biskyrmions~\cite{yu2014biskyrmion,peng2017real}, multi-$\pi$ skyrmions~\cite{bogdanov1999stability, zhang2016control,zheng2017direct,zhang2018real}, bobbers~\cite{rybakov2015new,zheng2018experimental}, and topologically trivial bubbles. Regarding applicability in spintronics the antiferromagnetic skyrmions~\cite{barker2016static,zhang2016magnetic,zhang2016antiferromagnetic,Bessarab2018, gobel2017afmskx,Shen2018,Akosa2018} that become antiferromagnetic bimerons (two mutually reversed bimerons on different sublattices), stand above all, since they allow for SOT-driven dynamics precisely in the middle of the racetrack at speeds of up to several $\mathrm{km}/\mathrm{s}$. Very recently the existence of such bimerons has been confirmed experimentally in synthetic antiferromagnets~\cite{Kolesnikov2018}.  
\begin{acknowledgments}
\change{
\paragraph{Note added.} After the submission of this Paper a vortex-antivortex crystal has been observed by Lorentz transmission electron microscopy in Co$_8$Zn$_9$Mn$_3$~\cite{yu2018transformation}. These textures have been predicted earlier~\cite{gaididei2012magnetic} and are topologically (but not geometrically) equivalent to bimeron crystals.}

We are grateful to Steffen Trimper and Stuart S. P. Parkin for fruitful discussions.
This work is supported by Priority Program SPP 1666 and SFB 762 of Deutsche Forschungsgemeinschaft (DFG). O.A.T. acknowledges support by the Grants-in-Aid for Scientific Research (Grant Nos. 17K05511 and 17H05173) from MEXT (Japan), by the grant of the Center for Science and Innovation in Spintronics (Core Research Cluster), Tohoku University, and by JSPS and RFBR under the Japan-Russian Research Cooperative Program.
\end{acknowledgments}
\bibliography{short,MyLibrary}
\bibliographystyle{apsrev}
\end{document}